\documentclass[preprintnumbers,secnumarabic,amsmath,amssymb,nofootinbib,floatfix]{revtex4}
\usepackage{graphics}
\usepackage{graphicx}
\usepackage{dcolumn}
\usepackage{bm}
\usepackage{mathrsfs}
\usepackage{pstricks}
\usepackage{color}
\usepackage{slashed}
\usepackage{amsmath}
\usepackage{epsfig}
\usepackage{amsfonts}
\usepackage{amssymb}

\def\beq{\begin{equation}}
\def\eeq{\end{equation}}
\def\bea{\begin{eqnarray}}
\def\eea{\end{eqnarray}}
\def\nn{\nonumber}
\begin{document}

\title{ Generalized unitarity method for unstable particles }
\author{Gabriel Menezes}
\email{gabrielmenezes@ufrrj.br}
\affiliation{~\\ Departamento de F\'{i}sica, Universidade Federal Rural do Rio de Janeiro, 23897-000, Serop\'{e}dica, RJ, Brazil \\ }
%

\begin{abstract}
In theories with unstable particles, unitarity is satisfied by the inclusion of only stable states in unitarity sums. Hence unitarity cuts are not to be taken through unstable particles. This raises a challenge to the generalized unitarity method, whose aim is to reconstruct amplitudes by analyzing sets of unitarity cuts. Nevertheless, under some general physical conditions, and perhaps some methodological modifications, we prove that the method is still reliable for one-loop amplitudes containing resonances. We discuss some simple examples which illustrate these features.
\end{abstract}

\maketitle

\section{Introduction}

Measurements of parameters of the Standard Model to very high precision is a landmark for the physics program of the Large Hadron Collider~\cite{LHC}. In recent years these have triggered an ongoing stream of research activities dedicated to assess precise predictions within perturbative quantum field theories. The immediate consequence is the development of modern tools to address the calculation of loop integrals. In this regard there is a pressing need to obtain a better knowledge of the analytic structure of such integrals to uncover more streamlined methods for their computation. This implies taking into account the investigation of the so-called cuts of internal propagators associated with intermediate particles in a given scattering amplitude. The major importance of cuts in this context is that it allows one to efficiently probe the analytic structure of loop integrals~\cite{Cutkosky:1960sp,smatrix}.

From unitarity constraints, we know that Feynman integrals should be multi-valued functions, whose discontinuities are precisely described by cuts. Actually, scattering amplitudes can be structured in terms of their singularities, so in principle the investigation of branch cuts and other singularities enables one to calculate loop amplitudes. {Four-dimensional} amplitudes that are uniquely specified by the nature of their branch cuts are said to be cut-constructible. Modern unitarity methods build on Landau conditions~\cite{Landau} in order to use cuts to set up projectors onto a basis of master integrals~\cite{Bern:1997sc,Britto:2004nc,Forde:2007mi,Kosower:12,Caron-Huot:2012awx,Johansson:2012zv,Johansson:2013sda,Abreu:2017ptx}. For recent applications, see Refs.~\cite{Sogaard:2014jla,Larsen:2015ped,Ita:2015tya,Remiddi:2016gno,Primo:2016ebd,Frellesvig:2017aai,Zeng:2017ipr,Dennen:2015bet,Dennen:2016mdk}. A precise definition of cuts is available for given classes of cuts. Among these, we can quote the so-called unitarity cuts which focus on a particular external channel~\cite{Veltmanbook,Remiddi:1981hn,Britto:2010xq,Abreu:2015zaa}. Historically, the unitarity method~\cite{Bern:1994zx,Bern:1994cg,Bern:1996je,Bern:2004ky,Britto:2008vq} was established as a systematic framework for one-loop evaluations, and is applicable to both supersymmetric and non-supersymmetric theories.

The standard practice of the unitarity method requires the replacement of two internal propagators by Dirac delta functions which projects the loop momenta they carry onto their on-shell values. On the other hand, in generalized unitarity~\cite{Bern:1997sc,Britto:2004nc,Bern:11,Bern:2010qa,Drummond:2008bq,Engelund:2013fja,Elvang:2019twd,Bern:2021ppb,Bern:2020ikv}, to be briefly reviewed below, one considers additional cut conditions to constrain other momenta to their associated on-shell values. As a consequence, if the momenta carried by more than two massless propagators take their on-shell values, the solutions to the cut conditions are now complex, which implies that the associated delta functions must give zero~\cite{Britto:2004nc,Kosower:12}. This observation has led to the concept that cuts should be computed via contour integration so that the associated contours should be suitably deformed in such a way as to encircle the poles of the cut propagators~\cite{Kosower:12,Cachazo:2008vp,Arkani-Hamed:2008owk}.

The obvious requirement here is that unitarity must be satisfied to all orders in perturbation theory. However, we know that many of our theories possess unstable particles which do not appear as asymptotic states. Should such unstable particles be incorporated nevertheless in unitarity relations? This issue was addressed by Veltman~\cite{Veltman:63,Diagrammar,Rodenburg,Lang-thesis,Denner:15}. The conclusion is that one should not take cuts through unstable propagators~\footnote{{The terminology ``unstable propagator" is actually a misnomer as what we mean by ``unstable" is the particle associated with this propagator. However, as the reader can easily check, this parlance is employed in the specialized literature. Therefore for brevity we will stick with this rather abuse of terminology. Henceforth, the expression ``unstable propagator" will be understood as ``propagator associated with an unstable particle/state".}}, hence unstable particles are not enclosed by unitarity sums. This can be generalized to encompass also unstable ghostlike resonances emerging in higher-derivative theories such as quadratic gravity and Lee-Wick theories~\cite{DM:19}. As the astute reader might have noticed, this might potentially put some obstructions to unitarity methods. Notwithstanding the foregoing remark, in this paper we will discuss how such methods are still solid for the calculation of loop amplitudes that comprise resonances of any type. Here we will use units such that $\hbar = c=1$. We take the Minkowski metric as $\eta_{\mu\nu} = \textrm{diag}(1,-1,-1,-1)$.

\section{Loop amplitudes and the unitarity method}

We are going to use tree-level amplitudes to reconstruct loop-level amplitudes. This is the so-called generalized unitarity method, a technique that we will now briefly describe{~\cite{Bern:1994zx,Bern:1994cg,Bern:1996je,Bern:2004ky,Britto:2008vq,Bern:11,Ellis:12,Frellesvig:Thesis,Brandhuber:05,Britto:11,Carrasco:11}}~\footnote{For a much more extensive body of research on the unitarity method, please check references within~\cite{Bern:11,Ellis:12,Frellesvig:Thesis,Brandhuber:05,Britto:11,Carrasco:11}. }.

The knowledge of tree amplitudes can be used to seek information about loop integrands. The action of taking loop propagators on-shell is known as a unitarity cut. It comes from the unitarity constraint of the S-matrix, which is a statement on the generalized optical theorem. That is, for an arbitrary process $a \to b$ one has that
\beq
i {\cal A}(a \to b) - i {\cal A}^{*}(b \to a)  =  - \sum_{f} \int d\Pi_{f} {\cal A}^{*}(b \to f) {\cal A}(a \to f)
(2\pi)^{4} \delta^{4}(a-f)
\label{opt_theo}
\eeq
where $d\Pi_{f}$ is the Lorentz-invariant phase space {measure}~\cite{Schwartz:13} and the sum runs over all possible sets $f$ of intermediate states and there is an overall delta function associated with energy-momentum conservation. In the above expression, {the ${\cal A}$'s} are (invariant) scattering matrix elements. {In a perturbative quantum field theory, when an  expansion in powers of a small coupling constant exists}, this constraint instructs us that the imaginary part of scattering amplitudes at a given order is obtained from the product of lower-order amplitudes. For instance, in the case of one-loop processes, one finds a product of two tree amplitudes on the right-hand side of Eq.~(\ref{opt_theo}). Usually this product presupposes the sum over all possible on-shell states that can cross the cut. One fundamental requirement is that {\it only states from the physical spectrum of the theory are allowed to be included in this sum}{~\cite{Veltman:63,DM:19}}. In a unitarity cut, we restrict the loop-momenta to be on-shell and only physical modes are enclosed in the two on-shell amplitudes on the right-hand side of Eq.~(\ref{opt_theo}). The cutting rules also consider integrals of any remaining freedom in the loop momentum after prescribing the so-called cut constraints (and, of course, momentum conservation). 

Unitarity cuts are efficient tools that enable one to relate the pole structure of the integrand with the branch-cut structure of the associated loop integral. Unitarity cuts can also involve more than two cut lines, which implies that several internal lines are taken on-shell. Here we say that we are able to reconstruct amplitudes from sets of generalized unitarity cuts. It turns out that such a set is overcomplete, which means that we can recourse to different strategies for extracting the relevant information. For example, one interesting approach is to use the method of maximal cuts{~\cite{Carrasco:11,Elvang:15}}. In this case we consider the maximum possible number of cut lines so that each cut furnishes a small piece of information. That is, we begin with generalized cuts possessing the maximum number of cut propagators. We use the information from such cuts to lay out an initial ansatz for the amplitude. Further cuts with reduced number of cut propagators are then considered and their information is systematically gathered in order to improve such an ansatz. The aim is to find an integrand that reproduces all the unitarity cuts. In principle this helps along the construction of the amplitude{~\cite{Carrasco:11,Ossola:2006us,Giele:2008ve,Ellis:2008ir}}.

{Here we will study generalized unitarity cuts on the level of the integrand, which can be written as a product of on-shell (tree-level or lower-loop) amplitudes. In particular, we are interested in considering maximal cuts consisting of only three-point tree amplitudes, namely:
\beq
\sum_{\textrm{states}} A^{\textrm{tree}}_{(1)} A^{\textrm{tree}}_{(2)} A^{\textrm{tree}}_{(3)} 
\cdots A^{\textrm{tree}}_{(m)} .
\label{2}
\eeq
}

The information from unitarity cuts can be used most efficiently if a complete basis of integrals is known. Indeed, all one-loop amplitudes in $D$ dimensions can be written as a sum of one-loop scalar integrals $I_{m}$, $m=1,2,3,\ldots,D$~\cite{Elvang:15}:
\beq
A_{n}^{1-\textrm{loop}} = \sum_{i} C_{D}^{(i)} I_{D;n}^{(i)}
+ \sum_{j} C_{D-1}^{(j)} I_{D-1;n}^{(j)} + \cdots
+ \sum_{k} C_{2}^{(k)} I_{2;n}^{(k)} + \sum_{l} C_{1}^{(l)} I_{1;n}^{(l)} 
+ \mathcal{R}
\label{oneloop}
\eeq
where $\mathcal{R}$ denotes rational terms (contributions that do not have branch cuts), $C_{D}^{(i)}$ are coefficients associated with tree-level amplitudes and $I_{m}^{(i)}$ are $m$-gon scalar integrals. In $D=4$, one-loop integrals reduce to a combination of box, triangle, bubble and tadpole scalar integrals~\cite{Britto:11,Henn:2014yza,vanNeerven:1983vr,Bern:1992em,Bern:1993kr,Brown:1952eu,thooft:79,Passarino:1978jh}. The latter are related to the coefficients $C_{1}^{(l)}$; such integrals vanish in dimensional regularization when only massless particles circulate in the loop. In $D=4$ power counting demonstrates that the {scalar box and triangle integrals} do not display UV divergences, but IR divergences due to possible massless corners. The bubble integrals have UV divergences but no IR divergences as both corners are massive. {Integrated results of such integrals can be found in several places in the literature, see for instance Refs.~\cite{Elvang:15,Bern:1992em,Bern:1993kr,thooft:79,Ellis:2007qk,Beenakker:1988jr,Denner:1991qq,Duplancic:2000sk}}.

{Therefore in four dimensions the following expansion is generically valid to any one-loop amplitude:
\beq
A^{1-\textrm{loop}} = \sum_{n=1}^{4} \sum_{\bf{K}} c_{n} (\bf{K}) I_{n}(\bf{K})
\label{exp1}
\eeq
where $K_i$ are sums of external momenta and $I_{n}$ are the associated scalar integrals. The coefficients $c_{n}$ are calculated using generalized cuts. For instance, consider a generic one-loop point amplitude written in the basis above. In this section we are working with only stable particles circulating in the loop. If we cut four propagators then the four dimensional integral becomes trivial:
\beq
\Delta_{4} A^{1-\textrm{loop}} = \int d^4 \ell \, G^{+}(\ell_{1}^2(\ell)) G^{+}(\ell_{2}^2(\ell)) 
G^{+}(\ell_{3}^2(\ell)) G^{+}(\ell_{4}^2(\ell))
A^{\textrm{tree}}_{1}(\ell) A^{\textrm{tree}}_{2}(\ell) A^{\textrm{tree}}_{3}(\ell) A^{\textrm{tree}}_{4}(\ell)
\label{cut4}
\eeq
where $A^{\textrm{tree}}_{j}$ are tree-level amplitudes and, using a spectral representation~\cite{DM:19}, 
\beq
G^{+}(p^2) = 2 \pi \theta(p_0) \int_{0}^{\infty} ds \, \delta(p^2 - s) \frac{\sigma(s)}{\pi}
\label{cut-stable}
\eeq
are the cut propagators (or positive-frequency Wightman functions) associated with stable particles. For brevity we have absorbed $2\pi$ factors into the definition of the loop integral in Eq.~(\ref{cut4}). Since in the case of stable particles the spectral function $\sigma(s)$ has a pole at one-particle states, we can also write that (assuming we are not above a given multi-particle threshold)
\beq
G^{+}(p^2) =  2 \pi \theta(p^0) \delta(p^2 - m^2).
\label{cut-stable2}
\eeq
In other words, the ``cut of a propagator" means removing its principal part while preserving the delta function imposing the on-shell condition.}

{For simplicity we have taken all internal propagators to have the same mass $m$, which can be zero. On the other hand, when applied to the master integrals, the quadruple cut selects the contribution from the box integral with momenta $K_1, K_2, K_3, K_4$ at the corners. Therefore
\beq
\Delta_{4} A^{1-\textrm{loop}} = c_{4}(K_1, K_2, K_3, K_4) \Delta_{4} I_{4}(K_1, K_2, K_3, K_4).
\eeq
where $I_{4}(K_1, K_2, K_3, K_4)$ is the associated $4$-point box scalar integral:
\beq
I_{4}(K_{1},K_{2},K_{3},K_{4}) = \int \frac{d^{D}\ell}{(2\pi)^D} 
\frac{1}{\ell^2 \left(\ell + K_{1}\right)^2 
\left(\ell + K_{1} + K_{2}\right)^2 \left(\ell - K_{4}\right)^2} .
\eeq
In particular, the quadruple cut of the scalar box integral is a Jacobian factor. This factor appears on both sides of the equation. Hence, comparing both expressions for $\Delta_{4} A^{1-\textrm{loop}}$, we see that the coefficient $c_{4}$ can be expressed as a product of tree-level amplitudes~\cite{Britto:11}: 
\beq
c_4 = \frac{1}{2} \sum_{\ell \in {\cal S}} \sum_{\textrm{states}}
A^{\textrm{tree}}_{1}(\ell) A^{\textrm{tree}}_{2}(\ell) A^{\textrm{tree}}_{3}(\ell) A^{\textrm{tree}}_{4}(\ell)
\label{c4}
\eeq
where the factor of $1/2$ emerges as there are exactly two solutions for the set ${\cal S}$ of cut conditions determined by the four delta functions of the cut propagators. Hence in principle the quadruple cut of the scalar box integral would suffice to calculate the box coefficient. Furthermore, this implies that the maximal cut in this case reads
\beq
A^{1-\textrm{loop integrand}}(\ell) \bigg|_{\textrm{Maximal cut}}
= \sum_{\textrm{states}}
A^{\textrm{tree}}_{1}(\ell) A^{\textrm{tree}}_{2}(\ell) A^{\textrm{tree}}_{3}(\ell) A^{\textrm{tree}}_{4}(\ell)
\eeq
which is a direct proof of the one-loop form of Eq.~(\ref{2}) for stable particles.}

{In conclusion, equipped with the integral reduction~(\ref{exp1}) valid to all one-loop amplitudes, and benefiting from the factorization property of the amplitude, by using unitarity methods
one is in a position to reconstruct one-loop amplitudes from tree-level information without the often burdensome Feynman diagram expansion. Moreover, we see that the application of the generalized unitarity method requires exploring further discontinuities which implies that a different number of propagators ought to be put on-shell in comparison with textbook unitarity cuts. And this can only be achieved if there is a contribution of an isolated simple pole at $p^2 = m^2$ (or $p^2 = 0$ for massless particles) coming from one-particle states -- in other words, if the cut propagators have the expected cut structure, as given by Eq.~(\ref{cut-stable2}).}

{Actually, one must be more careful when resorting to generalized unitarity, since the solutions to the cut conditions are generally complex, leading to delta functions that trivially yield zero. The solution is to use contour integration. That is, instead of replacing the propagators by delta functions, one must replace the original contour of integration~\cite{Kosower:12}. In summary, the idea is that, as the support of the delta functions is outside the physical region, the integration procedure is implemented in terms of contour integrals in $\mathbb{C}^4$, the loop momentum being regarded as a complex vector. Such contours are such that their product encircles the poles in the four-dimensional components of the loop momentum. By performing the four-dimensional loop-momentum integral over each contour, the residue at the corresponding encircled pole is attained. In fact, one defines the product of delta functions to generate {\it exactly} this contour integral~\cite{Kosower:12}.} 

{Of course, the aforementioned operation does not leave expression~(\ref{exp1}) intact, as there are terms that integrate to zero in the original contour which no longer necessarily vanish if we integrate over general contours in the complex plane. In order to do away with such spurious terms, one evaluates the integral over a suitable linear combination of new contours in such a way that such additional contributions are always projected out. This produces the coefficients of the box integrals as given in Eq~(\ref{c4}). For a careful survey of all subtleties associated with this discussion, see Ref.~\cite{Kosower:12}. See also Ref.~\cite{Abreu:2017ptx}.}

\section{Unitarity method for unstable particles}

\subsection{Possible issues with unstable particles}

{In order to understand what could be the general issues involving unstable particles, let us imagine a given reaction consisting of the scattering of $a$ particles
\beq
a_1 + a_2 + \cdots + a_n \to A + c_1 + c_2 + \cdots + c_n
\eeq
producing final products given by a collection of $c$ particles and a particle labeled $A$. If all final products are described by stable particles, then in principle there is no concern in evaluating an on-shell amplitude such as ${\cal A}(a \to A + c)$ in any given order in perturbation theory. However, if $A$ is heavy enough, its coupling to lighter states in the theory, say labeled by $b$, makes it decay:
\beq
A \to b_1 + b_2 + \cdots + b_n
\eeq
and now the scattering process we have to consider is $a \to b + c$ with the associated on-shell amplitude ${\cal A}(a \to b + c)$. In this case $A$ enters the calculation as a virtual particle, not as an external state, which implies, by the Feynman rules, the presence of its propagator $1/(p^2 - m_A^2)$ in internal lines of given Feynman diagrams. The point is, if the phase space contains the resonance region 
$(\sum p_b)^2 = m_A^2$, then the results calculated from perturbation theory cannot be trusted close to this region.}

{In other words, basically a diagram with a single internal $A$ propagator must met one of the following conditions~\cite{Rodenburg}: $(i) m_A < \sum m_b$, which always happen if $A$ is stable or if $A$ is an unstable particle with threshold $E < \sum m_b$ such that it cannot decay into $b$. In this case perturbation theory can still be trustworthy~\footnote{For a nice discussion of some additional subtleties that can be encountered in this situation, see Ref.~\cite{Rodenburg} and references cited therein.}; $(i) m_A > \sum m_b$, which implies that $A$ is unstable and can decay to $b$-particles. In this situation, the phase space contains the resonance, and perturbation theory can no longer be generically trusted.}   

{The most direct approach to solve this problem is to consider a resummed form for the propagator of the unstable particle. However, for gauge theories, the resummation procedure must be done carefully, otherwise one might expect to be confronted with issues associated with gauge invariance and gauge-fixing parameter dependence~\cite{Rodenburg,Lang-thesis,Denner:15,Papavassiliou:1995fq,Papavassiliou:1995gs,Papavassiliou:1996zn,Papavassiliou:1997fn,Papavassiliou:1997pb}. Another possible approach is provided by the so-called complex-mass scheme (CMS)~\cite{Denner:1999gp,Denner:2005fg,Denner:2006ic}. In few words, it corresponds to a suitable generalization of the on-shell renormalization scheme. In the latter, the renormalized mass $m$ is specified by demanding $p^2 = m^2$ to be the pole position associated with the resummed propagator. This is fine for stable particles -- for unstable particles, the self-energy acquires an imaginary part, and as a consequence the renormalized mass does not correspond to the pole position. The modification proposed by the complex-mass scheme is the following: Define a {\it complex} renormalized mass 
$\bar{m}$ by requiring that $p^2 = \bar{m}^2$ matches the pole position of the resummed propagator for unstable particles. The fact that $\bar{m}$ is complex, and therefore cannot be associated with a physical entity, should be of no concern as renormalized parameters in the Lagrangian do not carry any physical meaning~\cite{Rodenburg}. For a recent discussion on the definition of the mass and width of a normal unstable particle, see Ref.~\cite{Willenbrock:2022smq}. } 

{One can show that this modification put forward by the complex-mass scheme avoids the aforementioned issues appearing in gauge theories as it renders unnecessary the resummation of internal propagators~\cite{Rodenburg,Lang-thesis,Denner:15}. Indeed, the bare propagator of the unstable particle $A$ (or its scalar part) within this method acquires the form (in momentum space) 
\beq
D^{\textrm{CMS}}(p^2) = \frac{1}{p^2 - \bar{m}_A^2}.
\label{prop-cms}
\eeq
By writing $\bar{m}_A^2 = m_A^2 - i m_A \bar{\Gamma}$, where $m_A$ and $\bar{\Gamma}$ are real, one can prove that the above propagator can be envisaged as the resummed form of the following propagator in a scheme in which the renormalized mass is given by $m_A$~\cite{Rodenburg}:
\beq
\frac{1}{p^2 - m_A^2 + \Sigma(p^2)} .
\eeq
The self-energy obeys $\Sigma(\bar{m}_A^2) =  i m_A \bar{\Gamma}$. That is, the bare propagator within the complex-mass scheme is intrinsically resummed. In addition, notice that, as 
$m_A \bar{\Gamma}$ is evaluated throughout the renormalization procedure, we must envisage this quantity as a function of the coupling constant $\lambda$ describing the interaction between the unstable particle and the lighter states, $m_A \bar{\Gamma} \sim {\cal O} (\lambda)$. For more technical details concerning the complex-mass scheme, we refer the reader Refs.~\cite{Rodenburg,Lang-thesis,Denner:15}.}

{Given the result~(\ref{prop-cms}), one can be tempted to think that the complex-mass scheme allows for an adequate spectral representation for the propagator of the unstable particle such that an unambiguous one-particle state contribution can be identified. That this is not straightforward can be seen as follows. Within the complex-mass scheme, the positive-frequency Wightman function associated with the unstable particle in momentum space reads~\cite{Rodenburg,Lang-thesis,Denner:15}
\beq
D^{+, \textrm{CMS}}(p^2) = - \textrm{Im}\left[ \frac{1}{\bar{p}^0} \frac{1}{p^0 - \bar{p}^0} \right]
\eeq
where
\beq
( \bar{p}^0 )^2 = {\bf p}^2 + \bar{m}_A^2 . 
\eeq
Observe that the CMS cut propagator does not quite have the correct cut structure as given by Eq.~(\ref{cut-stable2}) and, as a result, in principle we cannot connect $D^{+, \textrm{CMS}}$ with physical particles carrying positive energy forward in time. However, at leading order $\bar{\Gamma}/\bar{m}_A \sim \lambda$ this is possible; when taking the limit $\bar{\Gamma} \to 0$, $D^{+, \textrm{CMS}}(p^2)$ turns into a nascent delta function,
\beq
D^{+, \textrm{CMS}}(p^2) \bigg|_{\bar{\Gamma} \to 0} \to  2 \pi \theta(p^0) \delta(p^2 - m_A^2)
\label{cut-prop-CMS}
\eeq
thereby recovering the standard cut structure for the cut propagator, which allows us to associate with the propagation of positive-energy physical particles. In general, we can write~\cite{Rodenburg,Lang-thesis,Denner:15}
\beq
D^{+, \textrm{CMS}}(p^2) = 
    \begin{cases}
      2 \pi \theta(p^0) \delta(p^2 - m_A^2) + {\cal O}(\lambda) & \text{near resonance,} \\
       F(p^2, \bar{\Gamma}/\bar{m}_A) & \text{off resonance,}
    \end{cases}
\label{cut-prop-CMS2}
\eeq
where the first equality is valid to leading order, and
\beq
F(p^2, \bar{\Gamma}/\bar{m}_A) = \sum_{n=1}^{\infty} a_{n}(p^2, \bar{\Gamma}/\bar{m}_A) 
\left( \frac{\bar{\Gamma}}{\bar{m}_A} \right)^{n} .
\eeq
So we see that $F(p^2, \bar{\Gamma}/\bar{m}_A)$ corresponds to higher-order contributions. For the phase space as a whole, the function $D^{+, \textrm{CMS}}(p^2)$ is suppressed as the imaginary part of 
$\bar{p}^0$ is small. However, in the region of resonance the small imaginary part yields a non-negligible contribution given by the nascent delta function above. This means that, outside the resonance region, where the CMS cut propagator does not have the correct cut structure (i.e., far from the poles of 
$D^{+, \textrm{CMS}}$), the cut of the CMS propagator of the unstable particle will produce a contribution of higher order in perturbation theory, which can thus be neglected. Only when one is close to the resonance region -- which can take place depending on the external momentum configuration of a diagram -- that the CMS cut propagator is non-negligible. These features persist when including corrections to the leading-order result~\cite{Rodenburg,Lang-thesis,Denner:15}. In any case, to leading order the cut of the unstable particle propagator is simply the cut through its one-loop correction; in other words, through stable particle propagators. This is a consequence of the fact that at one-loop order the widths in the Complex-Mass Scheme and the traditional on-shell scheme coincide. }

\subsection{When the unitarity method works for unstable particles}

{As shown in Refs.~\cite{Veltman:63,DM:19}, in a theory with unstable particles (of any kind), unitarity is satisfied by the sole inclusion of asymptotically stable states. This means that cuts should not be taken through the unstable particles. Unitarity-based methods represent a kind of generalization of the optical theorem in that they investigate discontinuities of an amplitude in several kinematical channels in order to fully reconstruct loop amplitudes. But if discontinuities of a given loop amplitude are given by the cutting rules, how can one make sense out of the method when one is to cut a propagator associated with an unstable particle?}

{Let us suppose that all internal propagators of a given one-loop amplitude describe unstable particles. Naively one would conclude that one cannot apply directly unitarity methods to unstable particles. Fortunately, this is not the end of the story. That the method can still be applied to these cases can be observed by recalling the aforementioned discussion of unstable particles within the complex-mass scheme. Indeed, as we have mentioned, at leading order the cut propagator reproduces the nascent delta function typical of stable particles when one is close to the resonance region. Therefore in the complex-mass scheme we write
\beq
\Delta_{4} A^{1-\textrm{loop}} = \int d^4 \ell \, D^{+, \textrm{CMS}}(\ell_{1}^2(\ell)) 
D^{+, \textrm{CMS}}(\ell_{2}^2(\ell)) 
D^{+, \textrm{CMS}}(\ell_{3}^2(\ell)) D^{+, \textrm{CMS}}(\ell_{4}^2(\ell))
A^{\textrm{tree}}_{1}(\ell) A^{\textrm{tree}}_{2}(\ell) A^{\textrm{tree}}_{3}(\ell) A^{\textrm{tree}}_{4}(\ell)
\label{cutCMS}
\eeq
so that, at leading order and close to the resonance region, we find that
\bea
\Delta_{4} A^{1-\textrm{loop}} \bigg|_{\bar{\Gamma} \to 0} &=& 
\int d^4 \ell \, \delta(\ell_{1}^2(\ell) - m^2) \theta(\ell^0_1)
\delta(\ell_{2}^2(\ell) - m^2) \theta(\ell^0_2) \delta(\ell_{3}^2(\ell) - m^2) \theta(\ell^0_3)
\delta(\ell_{4}^2(\ell) - m^2) \theta(\ell^0_4)
\nn\\
&\times&
A^{\textrm{tree}}_{1}(\ell) A^{\textrm{tree}}_{2}(\ell) A^{\textrm{tree}}_{3}(\ell) A^{\textrm{tree}}_{4}(\ell) .
\eea
This implies that the coefficient $c_{4}$ of the box integral is still given by Eq.~(\ref{c4}) at leading order. In particular, the maximal cut of the one-loop amplitude is also given by the one-loop form of Eq.~(\ref{2}), which represents a proof of this result at leading order to the case of unstable particles running in the loop. Moreover, it is also clear when this procedure cannot be trusted -- this is when is off resonance, so that we are not able to put the internal momenta on-shell, see Eq.~(\ref{cut-prop-CMS2}). So when cutting an internal line corresponding to an unstable particle off resonance, the result we obtain is not a contribution to the imaginary part of the scattering amplitude, and, as a consequence, not a valid contribution to the coefficients of the scalar integrals in the above expansion given by Eq.~(\ref{exp1}). So it is not at all clear whether generalized cuts of propagators associated with unstable particles produce sensible results in this case. We will get back to this off-resonance topic shortly.}

{At higher orders, as the cut of a propagator associated with unstable particles must correspond to the cut through a loop of stable particles when one is close to the resonance region, again we find the correct cut structure. As a result, we believe that Eq.~(\ref{2}) must still be valid to all orders in perturbation theory, but a general proof of this result is beyond the scope of the present work. This is an interesting exploration, and we hope to return to this calculation in the near future.} 

{This discussion shows us that the unitarity method still makes sense in the case of unstable particles running in the loops; in order to implement the technique in a straightforward way, one must ensure that external momentum configurations of an amplitude allows the unstable particle propagator to become resonant. In this case the unstable particle cut propagator will have the correct cut structure to guarantee that unitarity is satisfied. In turn, from previous discussions, we know that the correct strategy for the cut of several propagators is to interpret the corresponding loop integral as a contour integral in $\mathbb{C}^4$. Moreover, the determination of the contours is such that it must encircle one-particle poles so that we can define the result of integrating over the product of delta functions as given by this contour integral. This is a necessary requirement on the grounds of unitarity -- cuts are applied only on the stable particles of the theory so that the sum in Eq.~(\ref{c4}) is guaranteed to be over only asymptotically stable states. It is only in this case that one can assert that the result of the integration over the contour $|p^2 - \bar{m}_A^2| = \varepsilon$ that encircle $p^2 - \bar{m}_A^2 = 0$ will represent an on-shell particle carrying positive energy forward in time. However, it is also clear that when one is off resonance, then one is also away from the pole of the propagator; the associated contour integration must have a vanishing residue in this case. To get a finite result, one must be close to the resonance region; in this case, the operation described in the previous section will yield a well-defined residue.}

{On the other hand, there is also other situation that the method can be applied without further issues -- this is the so-called narrow-width approximation (NWA). In this situation, the coupling to the decay products is taken to be sufficiently small so that only resonance production is significant. In this limit, we can take
\beq
i D(q) = \frac{i}{q^2-m^2 + i \gamma}  \ \ .
\eeq
Here $\gamma = \Gamma m$, where $\Gamma$ is the width of the resonance. In the narrow-width approximation, $\Gamma \ll m$ and hence
\beq
\textrm{Im}[D(q)] \sim - \pi \delta(q^2-m^2),
\eeq
that is, near the resonance, we can treat the resonant particle as being on-shell. This means that in this limit the cut taken through the unstable particle with $\Gamma \to 0$ recovers the result from the cut through the decay products~\footnote{Incidentally, this implies that the expansion~(\ref{exp1}) is also valid in the NWA as $\Gamma \to 0$ and the full energy dependence of the self-energy does not need to be taken into account.}. So effectively the NWA allows us to regard a long-lived resonance as being approximately a stable particle. Moreover, for gauge theories, the NWA does not suffer either from the gauge invariance problem alluded to above~\cite{Rodenburg}. Therefore in this situation the usual reasoning that lies behind the generalized unitarity method can be fully applied.}  

{In other words, for unstable particles the present practice of the unitarity method is valid if the assumption of a resonant unstable propagator is warrant. This can happen depending on external momentum configurations (and this can be proved at least in the Complex-Mass Scheme, as discussed above) or else one should verify whether the narrow-width approximation holds in the particular case under studied. In the complex-mass scheme, for a resonant unstable propagator, one can show that the cut of this propagator follows through the cut of only stable particles, preserving unitarity in Veltman's sense (i.e., by using the Largest Time Equation and employing suitably defined cut propagators).  At higher-orders life will not be so simple, but in any case one can still prove that unitarity is satisfied.}

\subsection{Lee-Wick theories}

{Now let us discuss Lee-Wick-type theories~\cite{Lee:1969fy,Cutkosky:1969fq,Coleman:1969xz,Grinstein:08,Grinstein:2008bg}. As well known, these class of theories have a ghost mode which can be directly seen from the propagator, which has the form
\begin{equation}\label{partialfrac}
  \frac{1}{q^2 - \frac{q^4}{M^2}} =  \frac{1}{q^2} -  \frac{1}{q^2 - M^2} \ \ .
\end{equation}
The overall negative sign in the second term signals that this pole is ghost-like. However, the coupling to light particles of the theory makes the heavy ghost state unstable. As discussed above, this implies that generically the associated propagator must be resummed to ensure the validity of perturbation theory. A spectral representation of the corresponding cut propagator can be written as~\cite{DM:19}
\bea
\widetilde{D}_{\textrm{LW}}^{\pm}(p^2) = - 2\pi \theta(\pm p_0) \int_{0}^{\infty} ds \, \delta(p^2 - s) \frac{\widetilde{\rho}(s)}{\pi} .
\eea
Following the same reasoning employed for normal unstable particles, we find that the cut of internal Lee-Wick propagators of a given one-loop amplitude cannot produce in general a contribution to its discontinuity. However, recall that  the structure of a normal resonance propagator is given by
\beq
iD(q) = \frac{i}{q^2 - m^2 + \Sigma(q)}   \  \ 
\eeq
with $\textrm{Im}[\Sigma(q)] > 0$. Now if the Lagrangian gets modified with a $\Box^2 $ term, the propagator accordingly is changed to
\beq\label{generalcase}
iD(q) = \frac{i}{q^2 - m^2 + \Sigma(q)-q^4/\Lambda^2} .
\eeq
Setting $\Lambda \to \infty$, we obtain the normal resonance. However for large finite $\Lambda$
we find a heavy-mass resonance, when $q^2 \sim \Lambda^2 $. Near this resonance, the propagator behaves as
\beq
iD(q) \sim \frac{-i}{q^2-\Lambda^2 -i \gamma} .
\eeq
The residue at this pole is always negative. Furthermore, the sign of the width is always opposite from normal. That is, for both finite $m$ and $\Lambda$, we verify the appearance of resonances of both types in the same propagator. In both cases, the imaginary part of the self-energy arising from the coupling to stable states is the same; notwithstanding it manifests itself in distinct ways near the resonances. This means that ghost resonances also obey a unitarity relation as a consequence of the fact that normal resonances satisfy this constraint~\cite{DM:19}.} 

{The above discussion shows us how to implement unitarity-based methods to one-amplitudes involving unstable ghost modes. In the complex-mass scheme we write the Lee-Wick propagator as
\beq
D^{\textrm{CMS}}_{\textrm{LW}}(p^2) = - \frac{1}{p^2 - \bar{M}^2}
\label{prop-cms-LW}
\eeq
where $\bar{M}^2 = M^2 - i M \bar{\Gamma}$. On the other hand, as the aforementioned discussion indicates, normal resonances and ghost-like resonances have a similar structure~\cite{DM:19}
\beq\label{twosigns}
iD(q) \sim \frac{Zi}{q^2-m^2 +iZ\gamma}
\eeq
with $Z=+1$ for a normal resonance and $Z=-1$ for the ghost resonance. The imaginary part is $Z$-independent,
\beq
\textrm{Im}[D(q)] \sim \frac{-\gamma}{(q^2-m^2)^2 + \gamma^2}  \ \ .
\eeq
This implies that the CMS Lee-Wick cut propagator will have the same structure as the propagator associated with a normal unstable particle within the complex-mass scheme. In particular, close to the resonance region, at leading order it will have the form given by Eq.~(\ref{cut-prop-CMS}), producing the correct cut structure. Hence Eq.~(\ref{2}) is also valid for Lee-Wick theories when one is close to the resonance region.}

{We finally remark that again we can also resort to the NWA in order to apply unitarity techniques to one-loop amplitudes with unstable ghost modes running in the loop. Nevertheless, we emphasize that one must be very careful when dealing with ghost modes in the NWA; in order to reproduce correctly the cuts one must resort to a modification of the contour in performing the loop momentum integration, as originally discussed by Lee and Wick~\cite{DM:19}.}

\subsection{When the unitarity method does not seem to work for unstable particles}

{Suppose we wish to study a particular process $a + b \to c +d$ which takes place exclusively through loops of unstable particles (of any type) and let us assume that we are off resonance. As asserted above, when one is off resonance the cut through the propagator of an unstable particle will always violate the cut structure. That is, the cut of an unstable particle propagator off resonance yields a contribution of higher order in perturbation theory -- such cuts can surely be disregarded. Hence if we use the reasoning above, the cuts of internal unstable propagators will produce a vanishing contribution, resulting in a vanishing amplitude by employing  the current practice of the unitarity method to reconstruct it. This is obviously an unsatisfactory answer since we know that amplitudes can be built using standard Feynman rules and Feynman diagrams. So can we take this as an indication that the unitarity method cannot be trusted in this case, as it seems that the associated amplitude (or some of its contributions) could not be determined from the knowledge of its cuts?}

{There are ways to circumvent this issue. For instance, recall that, in a theory containing unstable particles, unitarity is satisfied by the inclusion of only stable states in unitarity sums. This suggests that, in order to generically implement the generalized unitarity method to a theory containing unstable particles, we must consider the inclusion of only cuts from stable states in unitarity sums. This means that in general one must be able to reformulate the theory in terms of the stable particles only, eliminating from the outset any unstable fields in the Lagrangian. But this will actually introduce non-local vertices in our description. There is one constraint that we should impose in this situation. In order to preserve unitarity, the only acceptable poles in tree-level amplitudes are the ones that come from propagators. Since non-local vertices may generate unphysical poles that would not be be consonant with an exchange of a physical particle, we must impose that such poles have zero residue. Or we must claim that the residues of all such spurious poles must cancel to give zero. Of course other constraints can also be imposed on the non-local vertices, such as proper infrared behavior, valid Ward identities, etc. For a recent interesting discussion of tree-level scattering amplitudes of a particular category of non-local field theories, see Ref.~\cite{Modesto:2021soh}. One-loop unitarity for a class of perturbative scalar quantum field theories with non-local operators of fractional order was established in Ref.~\cite{Calcagni:2021ljs}.}

\section{Examples of the use of the unitarity method for unstable particles}

We now proceed to discuss with some detail three examples which can be relevant for particle physics in order to see how one can implement the unitarity method when unstable particles run inside loops in scattering amplitudes.

\subsection{Normal unstable particles}

\begin{figure}[htb]
\begin{center}
\includegraphics[height=60mm,width=75mm]{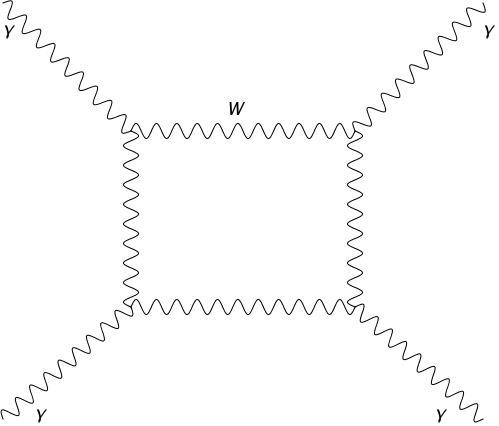}
\caption{ The box diagram related to the one-loop corrections to the process $\gamma\gamma \to \gamma\gamma$ via $W$-boson loops.}
\label{gammabox}
\end{center}
\end{figure}

Let us begin our discussions with normal unstable particles. Here we wish to investigate the {one-loop helicity amplitude $A^{1-\textrm{loop}}(++++)$ associated with} $\gamma-\gamma$ scattering via $W$ loops in the Standard Model~\cite{Boudjema:87,Jiang:93,Dong:93,Jikia:94,Yang:95}. It is known that the reaction $\gamma\gamma \to \gamma\gamma$ via $W$-boson at one-loop is finite~\cite{Fanchiotti:1972yf}. The standard Feynman-diagram formulation proceeds via box, triangle and bubble diagrams in which one must allow also for unphysical Higgs bosons (when working in a suitable non-linear $R_{\xi}$ gauge) and Faddeev-Popov ghosts in the loops (besides, of course, the $W$ particles). For our study we do not need to consider the unphysical particles; the finiteness of the amplitude will be easily established as we will see. We will use the method of maximal cuts in order to evaluate the amplitude in the expansion in terms of one-loop master integrals. The diagram is depicted in Fig.~\ref{gammabox}.

Since the $W$ boson is heavy, it is unstable and hence in principle we may not be allowed to cut the $W$ internal lines. However, the $W$ mass $M$ is about $80$ GeV and its decay width 
$\Gamma$ is about $2$ GeV~\cite{ParticleDataGroup:2020ssz}, hence $\Gamma \ll M$ and in principle one is justified in resorting to the narrow-width approximation, at least in a primary analysis. In this case, the production and decay of the resonance can be treated approximately in a separate way. {As discussed above, the propagator in the NWA has the correct cut structure and hence one can safely use the unitarity method in order to reconstruct the aforementioned box amplitude. On the other hand, as mentioned all one-loop integrals can be written in terms of a basis of scalar one-loop integrals as in Eq.~(\ref{oneloop}), so we are in safe ground here -- we can trust the results obtained here. In any case, we can also contemplate our results as an independent check of the helicity amplitude calculated in Ref.~\cite{Jikia:94}; see also Ref.~\cite{Costantini:1971cj}. }

Using the method of maximal cuts, in the narrow-width approximation the coefficient of the scalar box function may be calculated from the formula
\beq
c_{4}(p_1, p_2, p_3, p_4) = \frac{1}{2}  
\sum_{\ell \in {\cal S}} \sum_{\textrm{pol.}}
A^{\textrm{tree}}_{1}(p_1, - \ell^{KL}_2, \ell^{IJ}_1) A^{\textrm{tree}}_{2}(p_2, -\ell^{MN}_3, \ell^{KL}_2) 
A^{\textrm{tree}}_{3}(p_3, -\ell^{PQ}_4, \ell^{MN}_3) A^{\textrm{tree}}_{4}(p_4, -\ell^{IJ}_1, \ell^{PQ}_4) .
\eeq
The above on-shell tree amplitudes involve {one} photon and two $W$ bosons. The latter carry explicit $SU(2)$ little-group indices associated with massive spinors. A review of the formalism designed to deal with massive particles can be found in the Appendix. In addition, ${\cal S}$ is the solution set for the four delta functions of the cut propagators:
$$
{\cal S} = \{ \ell | \ell_1^2 = \ell^2 = M^2, \ell_2^2 = (\ell + p_1)^2 = M^2, 
\ell_3^2 = (\ell + p_1 + p_2)^2 = M^2, \ell_4^2 = (\ell_3 + p_3)^2 = (\ell - p_4)^2 = M^2 \} 
$$
where we took all external momenta incoming. Notice that the cut conditions imply that
\bea
2 p_{1} \cdot \ell &=& (\ell + p_1)^2 - \ell^2 - p_{1}^2 = 0
\nn\\
2 p_{3} \cdot \ell_3 &=& (\ell_3 + p_3)^2 - \ell_{3}^2 - p_{3}^2 = 0 .
\eea
Let us calculate the $3$-particle amplitude that appears above. Feynman rules will tell us that
\bea
i A_{3}(p^{s}_1, k^{KL}_{2}, k^{IJ}_1) &=& i e \bigl[ (k_1 - k_2)_{\rho} \eta_{\mu\nu} 
+ (k_2 - p_1)_{\mu} \eta_{\nu\rho} + (p_1 - k_1)_{\nu} \eta_{\rho\mu} \bigr] 
\epsilon^{\mu IJ}(k_1) \epsilon^{\nu KL}(k_2) \epsilon^{\rho s}(p_1;r) 
\nn\\
&=&  i e \biggl\{ \Bigl[ (k_1 - k_2) \cdot \epsilon^{s}(p_1;r) \Bigr] 
\Bigl[ \epsilon^{IJ}(k_1) \cdot \epsilon^{KL}(k_2) \Bigr]
+ \Bigl[ (k_2 - p_1) \cdot \epsilon^{IJ}(k_1) \Bigr] \Bigl[ \epsilon^{KL}(k_2) \cdot \epsilon^{s}(p_1;r) \Bigr]
\nn\\
&+& \Bigl[ (p_1 - k_1) \cdot \epsilon^{KL}(k_2) \Bigr] 
\Bigl[ \epsilon^{IJ}(k_1) \cdot \epsilon^{s}(p_1;r) \Bigr] \biggr\} .
\eea
Let us first choose $s=+$. Resorting to a bold notation for massive spinors, one finds
\bea
A_{3}(p^{+}_1, {\bf k}_{2}, {\bf k}_1) &=& 
\frac{e}{\sqrt{2} M^2 \langle r1 \rangle} \biggl\{ - \Bigl( \langle k_{1A} r \rangle 
\bigl[ 1 k^{A}_1\bigr]
-  \langle k_{2A} r \rangle \bigl[ 1 k^{A}_2 \bigr] \Bigr) \langle {\bf k}_1 {\bf k}_2 \rangle 
\bigl[ {\bf k}_2 {\bf k}_1 \bigr]
\nn\\
&+& \Bigl(  \langle {\bf k}_1 k^A_2 \rangle \bigl[ k_{2A} {\bf k}_1 \bigr] 
- \langle {\bf k}_1 1 \rangle \bigl[ 1 {\bf k}_1 \bigr] \Bigr)
 \langle {\bf k}_2 r \rangle \bigl[ 1 {\bf k}_2 \bigr] 
\nn\\
&+& \Bigl( \langle {\bf k}_2 1 \rangle \bigl[ 1 {\bf k}_2 \bigr] 
- \langle {\bf k}_2 k^A_1 \rangle \bigl[ k_{1A} {\bf k}_2 \bigr] \Bigr)
 \langle {\bf k}_1 r \rangle \bigl[ 1 {\bf k}_1 \bigr] \biggr \}
\nn\\
&=& - \frac{\sqrt{2} e}{ M^2}  \frac{\langle r| {\bf k}_1 | 1\bigr]}{\langle r1 \rangle} 
\langle {\bf k}_1 {\bf k}_2 \rangle^2  
\eea
where we used the Schouten identities and the following relations (which follow from the Schouten identities)~\cite{Durieux:20}
\bea
\langle r| {\bf k}_1 | 1\bigr] \langle {\bf k}_1 {\bf k}_2 \rangle
&=&  M \Bigl( \langle r {\bf k}_{1} \rangle \bigl[ 1 {\bf k}_2  \bigr] 
+  \langle r {\bf k}_{2} \rangle \bigl[ 1 {\bf k}_1 \bigr]  \Bigr)
\nn\\
\langle r| {\bf k}_1 | 1\bigr] \bigl[ {\bf k}_1 {\bf k}_2 \bigr]
&=&  M \Bigl( \langle r {\bf k}_{1} \rangle \bigl[ 1 {\bf k}_2  \bigr] 
+  \langle r {\bf k}_{2} \rangle \bigl[ 1 {\bf k}_1 \bigr]  \Bigr)
+ \langle  r1 \rangle 
\bigl[ 1 {\bf k}_1 \bigr]  \bigl[ 1 {\bf k}_2 \bigr] .
\eea
With an almost identical calculation for the $s=-$ case, one finds that
\beq
A_{3}(p^{-}_1, {\bf k}_{2}, {\bf k}_1) = 
 - \frac{\sqrt{2} e}{ M^2}  \frac{\bigl[ r| {\bf k}_1 | 1\rangle}{\bigl[ 1r \bigr]} 
\bigl[ {\bf k}_1 {\bf k}_2 \bigr]^2  .
\eeq
For concreteness, let us choose a specific set of helicities for the external photons. Now the cut reads
(sum over repeated $SU(2)$ little-group indices is implicit)
\bea
\hspace{-7mm}
A^{1-\textrm{loop integrand}}_{4}(\ell) \bigg|_{\textrm{quadruple cut}} &=&
A^{\textrm{tree}}_{1}(p^{+}_1, - \ell_{2KL}, \ell^{IJ}_1) 
A^{\textrm{tree}}_{2}(p^{+}_2, -\ell_{3MN}, \ell^{KL}_2) 
A^{\textrm{tree}}_{3}(p^{+}_3, -\ell_{4PQ}, \ell^{MN}_3) 
A^{\textrm{tree}}_{4}(p^{+}_4, -\ell_{1IJ}, \ell^{PQ}_4)
\nn\\
&=& \frac{4 e^4}{ M^8}
 \frac{\langle r_1| \boldsymbol\ell_1 | 1\bigr]}{\langle r_1 1 \rangle} 
\langle \boldsymbol\ell_1 \boldsymbol\ell_2 \rangle^2
 \frac{\langle r_2| \boldsymbol\ell_2 | 2\bigr]}{\langle r_2 2  \rangle} 
\langle \boldsymbol\ell_2 \boldsymbol\ell_3 \rangle^2
\frac{\langle r_3| \boldsymbol\ell_3 | 3\bigr]}{\langle r_3 3 \rangle} 
\langle \boldsymbol\ell_3 \boldsymbol\ell_4 \rangle^2
\frac{\langle r_4| \boldsymbol\ell_4 | 4\bigr]}{\langle r_4 4 \rangle} 
\langle \boldsymbol\ell_4 \boldsymbol\ell_1 \rangle^2 
\nn\\
&=& - \frac{4 e^4}{ M^4} \frac{ s_{12} s_{23} }
{\langle 1 2 \rangle \langle 23 \rangle \langle 3 4 \rangle \langle 41 \rangle} 
M^6 \langle \boldsymbol\ell_4 \boldsymbol\ell_4 \rangle^2 
= - \frac{4 e^4}{ M^4} \frac{ s_{12} s_{23} }
{\langle 1 2 \rangle \langle 23 \rangle \langle 3 4 \rangle \langle 41 \rangle} 
M^8 \Bigl( \delta^{P}_{P} \delta^{Q}_{Q} + \delta^{P}_{Q} \delta^{Q}_{P}  \Bigr)
\nn\\
&=& - 24 e^4 M^4 \frac{ s_{12} s_{23} }
{\langle 1 2 \rangle \langle 23 \rangle \langle 34 \rangle \langle 41 \rangle} 
\eea
where we used the on-shell conditions, symmetrization of $SU(2)$ little-group indices and also that 
$|\lambda^{I} \rangle^{\alpha} \langle \lambda_{I} |_{\beta} = -M \delta^{\alpha}_{\beta}$ and $\langle \lambda^{I} \lambda_{J} \rangle = M \delta^{I}_{J}$. In addition, we have chosen $r_1 = p_2$,  $r_2 = p_1$, $r_3 = p_4$ and $r_4 = p_3$ and, as usual, 
$s_{ij} = (p_i + p_j)^2$. The convention we use here is the following:
\bea
|-p\rangle &=& - |p\rangle, \,\,\, \langle-p| = -\langle p|
\nn\\
|-p\bigr] &=& |p\bigr], \,\,\, \bigl[-p| = \bigl[p| .
\eea
We have so far calculated the coefficient associated with the scalar box integral. In order to calculate the coefficients of triangles, bubbles and tadpoles, one must resort to lower-order cuts. For instance, a triple-cut reads 
\beq
A^{1-\textrm{loop integrand}}_{4}(\ell) \bigg|_{\textrm{triple cut}} = 
A^{\textrm{tree}}_{3}(p^{+}_1, - \ell_{2KL}, \ell^{IJ}_1) 
A^{\textrm{tree}}_{3}(p^{+}_2, -\ell_{3MN}, \ell^{KL}_2) 
A^{\textrm{tree}}_{4}(p^{+}_4, -\ell_{1IJ}, \ell^{MN}_3, p^{+}_3).
\eeq
The other possible three-particle cut diagrams are obtained from this one by cyclic relabeling of the external particles. The cut conditions are given by
\bea
\ell_1^2 &=& \ell^2 = M^2 
\nn\\
\ell_2^2 &=& (\ell + p_{1})^2 = M^2
\nn\\ 
\ell_3^2 &=& (\ell + p_1 + p_2)^2 = M^2 . 
\eea
Notice that these imply that $\ell \cdot p_1 = 0$. One finds that
\beq
A^{1-\textrm{loop integrand}}_{4}(\ell) \bigg|_{\textrm{triple cut}} = 
- 24 e^4 M^{4} \frac{ s_{12} s_{23} }
{\langle 1 2 \rangle \langle 23 \rangle \langle 34 \rangle \langle 41 \rangle}
\frac{1}{(\ell - p_4)^2 - M^2}
\eeq
where we have chosen $r_1 = 2$ and $r_2 = 1$ and we used the cut conditions. As for the two-particle cut, we find
\beq
A^{1-\textrm{loop integrand}}_{4}(\ell) \bigg|_{\textrm{double cut}} = 
{A^{\textrm{tree}}_{4}(p^{+}_2, -\ell_{3MN}, \ell^{IJ}_1,p^{+}_1)} 
A^{\textrm{tree}}_{4}(p^{+}_4, -\ell_{1IJ}, \ell^{MN}_3, p^{+}_3).
\eeq
As above, the other possible two-particle cut diagrams are obtained from this one by cyclic relabeling of the external particles. The cut conditions are given by
\bea
\ell_1^2 &=& \ell^2 = M^2 
\nn\\
\ell_3^2 &=& (\ell + p_{1} + p_{2})^2 = M^2 . 
\eea
One finds that
\beq
A^{1-\textrm{loop integrand}}_{4}(\ell) \bigg|_{\textrm{double cut}} 
= - 24 e^4 M^{4}
\frac{ s_{12} s_{23} }{\langle 1 2 \rangle \langle 23 \rangle \langle 34 \rangle \langle 41 \rangle}
\frac{1}{(\ell+p_1)^2 - M^2}
\frac{1}{(\ell - p_4)^2 - M^2} 
\eeq
where we used momentum conservation and the cut conditions.
 
Now let us discuss our results. Concerning the triple cut, there are two possible integrals that can contribute, namely the box integral and the triangle integral. However, our result shows the presence of one uncut propagator. So this would exclude triangle integrals from the expansion. To confirm this, let us analyze the $2$-particle cut. Again box and triangle integrals contribute, and now also bubble integrals can contribute. Nevertheless, our result shows the presence of two uncut propagators. This confirms the exclusion of triangle integrals from the expansion, and also states the absence of bubble integrals. We can perform a single cut to confirm that there will remain three uncut propagators in the result. Hence, the final answer is that only the box integral is present. So finally we can write
\beq
A^{1-\textrm{loop}}_{4}(++++) = - 24 e^4 M^4 \frac{ s_{12} s_{23} }
{\langle 1 2 \rangle \langle 23 \rangle \langle 34 \rangle \langle 41 \rangle}
I_{4}(p_{1},p_{2},p_{3},p_{4}) + \textrm{Perm.}  + {\cal R}
\eeq
where {$\textrm{Perm.}$ indicates permutations of external particles}, ${\cal R}$ comprise rational terms and 
\bea
I_{4}(p_{1},p_{2},p_{3},p_{4}) &=& \int \frac{d^{D}\ell}{(2\pi)^D} 
\frac{i}{[\ell^2 - M^2 + i M \Gamma]} \frac{i}{[ \left(\ell + p_{1}\right)^2 - M^2 + i M \Gamma]}
\nn\\
&\times& \frac{i}{[\left(\ell + p_{1} + p_{2}\right)^2 - M^2 + i M \Gamma]} 
\frac{i}{[\left(\ell - p_{4}\right)^2 -M^2 + i M \Gamma]} 
\eea
with, as already quoted, $\Gamma \ll M$. We can think of the presence of $\Gamma$ in the above equation as a consequence of the fact that, for unstable particles, we should use a resummed form for its propagator. 
{In any case, one should bear in mind that, as we are considering the decay width to be very small, one must envisage $I_{4}(p_{1},p_{2},p_{3},p_{4})$ in the limit $\Gamma \to 0$, which ought to be taken at the end of the calculations. Otherwise, one can prove that the coefficient of the box will also display a finite $\Gamma$-dependence which is not captured by the unitarity method. As well known, the inclusion of the appropriate dependence on $\Gamma$ both in the propagators as well as in the corresponding coefficients of the integral is mandatory to ensure the correct gauge cancellations. Nevertheless, as we are considering in this calculation a situation which is dominated by production of on-shell unstable particles with a vanishingly small decay width, finite-width effects are negligible as long as the required precision is taken to be small in comparison with $\Gamma/M$~\footnote{{Recall this is the narrow-width approximation we are using, not a fixed-width scheme, which is known to introduce gauge dependence. Indeed, fixed-width schemes are known to violate SU(2) $\times$ U(1) Ward identities -- see for instance the discussion in Ref.~\cite{Argyres:1995ym}. Perhaps, in order to avoid further confusion, a better terminology for the narrow-width approximation would be ``zero-width approximation", as first suggested by John F. Donoghue in a private communication with the author, because after all it comes from taking the width to zero.}}.}

Rational terms are not detected by unitarity cuts. Hence the above result have potential ambiguities in rational functions. In order to remove such ambiguities one may consider dimensionally regularized representations for the tree amplitudes. This means considering $d$-dimensional cuts, with $d = (4-2\epsilon)$. Photons live in four dimensions, whereas the loop momentum is $d$-dimensional. In this case one has to be careful when dealing with the summation over states.

A crucial component of generalized unitarity cuts is the sum over physical states. One must be careful in the sum over the physical states of gauge bosons in $d$ dimensions~\cite{Bern:19}. It is given by the so-called physical state projector:
\beq
P_{\mu\nu} = \sum_{\textrm{pols.}} \epsilon_{\mu}(-p) \epsilon_{\nu}(p) = - \eta_{\mu\nu} + \cdots
\eeq
where the ellipsis stand for terms depending on an arbitrary null reference momentum (for massless particles) or on the mass of the particle (for massive particles). In the present case, we will only be concerned with the maximal cut since we already know that only the box integral is present.
{Here we simply adopt the four-dimensional helicity scheme~\cite{Bern:1991aq,Bern:96} in which all internal and external states (and also polarization vectors) are four-dimensional and loop momentum and phase-space integrals are in $d = 4-2\epsilon$ dimensions. There are no remaining ambiguities to be considered in our case as the amplitude under consideration vanishes at tree level and there are no ultraviolet divergences.}

In general, in the evaluation of the quadruple-cut, we have to discriminate between the dimension of loop momenta and the dimension of the space of physical states; in other words, we should envisage any factor of $D$ emerging from contracting Lorentz indices ($\delta^{\mu}_{\mu} = D$) as a different quantity in comparison with the dimension $d$ of the loop momenta for which we take $d = 4-2\epsilon$. In the limit that $D \to d$ one should obtain the same result as before, except that the mass has undergone the shift $M^2 \to M^2 + \mu^2$, where $\mu^{\alpha}$ is a vector associated with the $(-2\epsilon)-$dimensional part of the loop momentum. This means that we should use the following modified cut conditions
\bea
\ell_1^2 &=& \ell^2 = M^2 + \mu^2
\nn\\
\ell_2^2 &=& (\ell + p_{1})^2 = M^2 + \mu^2
\nn\\ 
\ell_3^2 &=& (\ell + p_1 + p_2)^2 = M^2 + \mu^2 
\nn\\
\ell_4^2 &=& (\ell_3 + p_3)^2 = (\ell - p_4)^2 = M^2 + \mu^2 . 
\eea
The final result is given by
\bea
A^{1-\textrm{loop}}_{4}(1^{+}, 2^{+}, 3^{+}, 4^{+}) &=&
- 24 e^4 \frac{i}{(4\pi)^{2-\epsilon}} \frac{ s_{12} s_{23} }
{\langle 1 2 \rangle \langle 23 \rangle \langle 34 \rangle \langle 41 \rangle}
 {\cal I}_{4}[(M^2 + \mu^2)^2] + \textrm{Perm.}
\eea
where~\cite{Brandhuber:05,Bern:96,Bern:97}
\bea
{\cal I}^{d}_{n}[\mu^{2r}] &=& i (-1)^{n+1} (4\pi)^{d/2} \int \frac{d^{-2\epsilon}\mu}{(2\pi)^{-2\epsilon}}
\int \frac{d^4 \ell}{(2\pi)^4} \frac{\mu^{2r}}{(\ell^2 - M^2 - \mu^2)((\ell+p_1)^2 - M^2 - \mu^2) \cdots 
((\ell+\sum_{i=1}^{n-1} p_i)^2 - M^2 - \mu^2)}
\nn\\
&=& - \epsilon(1-\epsilon) \cdots (r-1-\epsilon) {\cal I}^{d+2r}_{n}[1]
\nn\\
{\cal I}^{d}_{n}[1] &=& i (-1)^{n+1} (4\pi)^{d/2} \int \frac{d^{-2\epsilon}\mu}{(2\pi)^{-2\epsilon}}
\int \frac{d^4 \ell}{(2\pi)^4} \frac{1}{(\ell^2 - M^2 - \mu^2)((\ell+p_1)^2 - M^2 - \mu^2) \cdots 
((\ell+\sum_{i=1}^{n-1} p_i)^2 - M^2 - \mu^2)}
\nn\\
\eea
and hence
\bea
{\cal I}_{4}[(M^2 + \mu^2)^2] &=& M^4 {\cal I}^{d=4-2\epsilon}_{4}[1] 
- 2 M^2 \epsilon {\cal I}^{d=6-2\epsilon}_{4}[1]
- \epsilon(1-\epsilon) {\cal I}^{d=8-2\epsilon}_{4}[1] 
\nn\\
- \epsilon {\cal I}^{d=6-2\epsilon}_{4}[1] &=& 0 + {\cal O}(\epsilon)
\nn\\
- \epsilon(1-\epsilon) {\cal I}^{d=8-2\epsilon}_{4}[1] &=& - \frac{1}{6} + {\cal O}(\epsilon).
\eea
We recall that for a complete removal of the ambiguity associated with the rational terms, additional procedures should be carried out~\cite{Bern:96}; however, such procedures are trivial in the present case since the associated tree-level amplitude vanishes and there are no ultraviolet divergences. Scalar box integrals were explicitly calculated in Refs.~\cite{Bern:96,thooft:79,Denner:1991qq}. Furthermore, as promised the helicity amplitude is free from UV divergences. Finally, by exploring the fact that
$$
\left| \frac{ -i s_{12} s_{23} }
{\langle 1 2 \rangle \langle 23 \rangle \langle 34 \rangle \langle 41 \rangle}  \right|^2 = 1
$$
and taking into account the different permutations over the external photons, one can easily see that our result agrees perfectly with the ones given in the literature~\cite{Jikia:94,Costantini:1971cj}, apart from an overall phase factor (which is unimportant); {one simply needs} to be careful with the different conventions on external momenta.

\subsection{Lee-Wick QED}

Now we will discuss a simple example coming from higher-derivative QED. The Lagrangian for the gauge sector reads~\cite{Grinstein:08}
{
\beq
 {\cal L} = - \frac{1}{4} F_{\mu\nu} F^{\mu\nu} 
+ \frac{1}{2 M^2} \partial_{\mu} F^{\mu\nu} \partial_{\lambda} F^{\lambda}_{\ \ \nu}.
\eeq
}
As well known, Lee-Wick Lagrangians can be rewritten by introducing auxiliary gauge bosons with a very large mass $M$, very much larger than any other particle masses in our problem. As extensively discussed elsewhere, the coupling of these auxiliary massive gauge bosons to light fields makes them decay, and positive energy is required to excite this resonance{~\cite{Donoghue:2021eto,DM:19}}. Furthermore, this resonance has a ``backwards in time" feature in that the propagator has the approximate form (close to the resonance)
\beq
iD(q) \sim \frac{-i}{q^2 -\bar{M}^2 -i \gamma}  
\eeq
where we have suppressed Lorentz indices. Notice that there are two minus sign differences from a normal resonance, the $-i$ in the numerator and the $-i\gamma$ in the denominator. These combined sign differences lead to the distinguishing property of a time-reversed version of a usual unstable particle propagator. This unusual resonance was dubbed a {\it Merlin mode} in Refs.~\cite{DM:19,Donoghue:2019ecz}. There are evidences that point to the stability of theories containing Merlin modes~\cite{DM:19,Donoghue:2021eto}.

\begin{figure}[htb]
\begin{center}
\includegraphics[height=60mm,width=70mm]{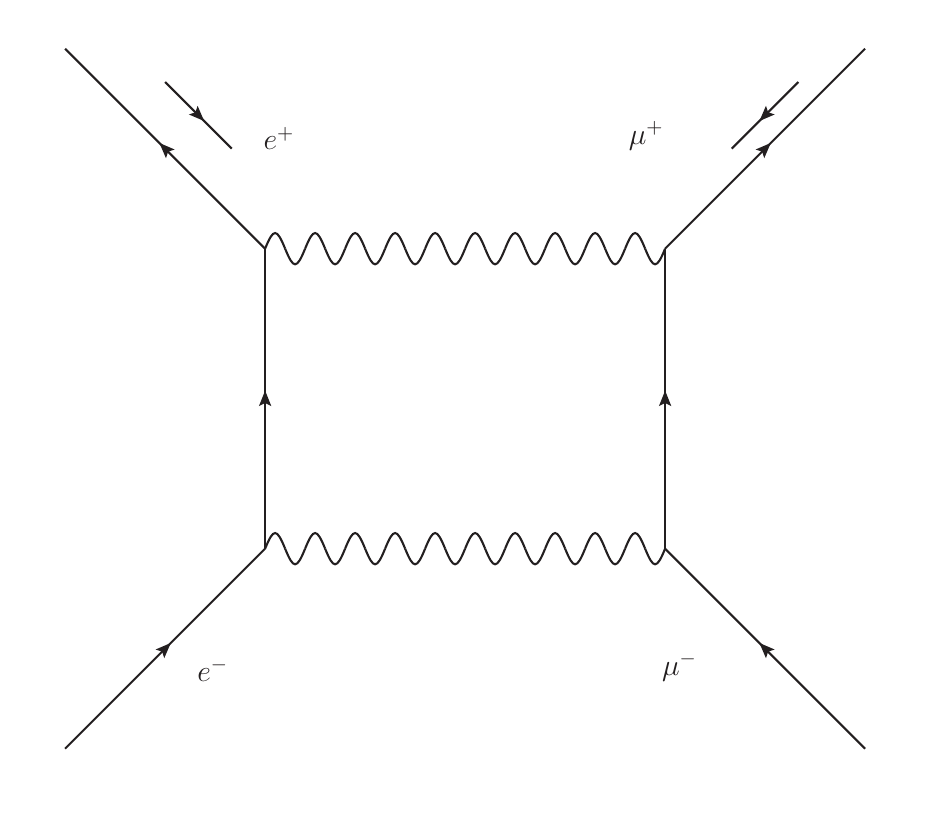}
\caption{ The box diagram related to the one-loop corrections to the process $e^+ e^- \to \mu^+ \mu^-$. Curly lines represent photons or Merlin particles and straight lines are fermions}
\label{photonboxmuon}
\end{center}
\end{figure}

Here we are interested in using the unitarity method to calculate a scattering amplitude involving Merlin particles circulating in the loop. For simplicity, we will consider the narrow-width approximation, 
$\Gamma \ll M$, where $\Gamma$ is the width of the Merlin particle. The process we have in mind is the muon-electron scattering $e^+ e^- \to \mu^+ \mu^-$ at next-to-leading order, which is one of the simplest in QED processes, but a crucial one in the comprehension of all reactions in $e^+ e^-$ colliders~\cite{Peskin:1995ev}. The calculations with photons running in the loop {were} carried out in a number of places, see for instance Refs.~\cite{Nikishov:60,Eriksson:61,Eriksson:63,Nieuwenhuizen:71,D'Ambrosio:83,Kukhto:87,Bardin:97,Kaiser:10,Alacevich:19}. Here we wish to consider solely the one-loop box diagram depicted in Fig.~\ref{photonboxmuon}. We show how to calculate the coefficient associated with scalar boxes when internal gauge lines are associated with Merlin propagators.

For simplicity, consider the high-energy scattering limit in which the fermions are massless. The $3$-particle amplitude involving a fermion, an anti-fermion and a Merlin particle reads (all momenta incoming)
\bea
i A_{3}(\bar{f}_1^{-1/2},f_2^{+1/2},{\bf p}) &=& 
\bar{v}_{-}(1) i e \gamma^{\mu} u_{+}(2) \epsilon^{IJ}_{\mu}
\nn\\
&=&  i e \langle 1| \gamma^{\mu} |2 \bigr] 
\frac{1}{\sqrt{2}} \, \frac{ \langle {\bf p} | \sigma_{\mu} | {\bf p} \bigr]  }{M}
\nn\\
&=&  \sqrt{2} i e \frac{\langle 1 {\bf p} \rangle \bigl[ {\bf p} 2 \bigr]}{M}
\eea
where we used that $\eta^{\mu\nu} \sigma_{\mu}^{\alpha \dot{\alpha}} \sigma_{\nu}^{\beta \dot{\beta}}
= 2 \varepsilon^{\alpha\beta} \varepsilon^{\dot{\alpha}\dot{\beta}}$. Notice that both fermions need to have opposite helicity to give a non-vanishing result. Since $\langle p| \gamma^{\mu} |q \bigr] = \bigr[q| \gamma^{\mu} |p\rangle$, one also obtains
\beq
A_{3}(\bar{f}_1^{+1/2},f_2^{-1/2},{\bf p}) = \sqrt{2} e \frac{\langle 2 {\bf p} \rangle \bigl[ {\bf p} 1 \bigr]}{M} .
\eeq
The $3$-point amplitudes involving fermions and a Merlin particle can be computed similarly.

In order to calculate the associated contribution to the coefficient of the scalar box integral, we will resort to the maximal-cut technique, which in the present case means evaluating a quadruple cut. We choose the associated helicities to be $h_{e^-} = h_{\mu^-} = -1/2$ and $h_{e^+} = h_{\mu^+} = + 1/2$. We have that
\beq
\textrm{Cut-max} =  \sum_{\{K\},\{L\}} 
A_{3}(2^{\prime},\ell_4^{-1/2}, - \ell_{1 \{K\}})
A_{3}(1^{\prime}, \ell_2^{-1/2},  \ell_1^{\{K\}}) 
A_{3}(- \ell_2^{1/2}, 1, \ell_3^{\{L\}}) 
A_{3}(- \ell_{4}^{1/2}, 2, - \ell_{3 \{L\}}) 
\eeq
with the following cut conditions:
\bea
\ell_1^2 &=& \ell^2 = M^2
\nn\\
\ell_2^2 &=& (- \ell - p^{\prime}_{1})^2 = 0
\nn\\ 
\ell_3^2 &=& (\ell_2 - p_1)^2 = (- \ell - p^{\prime}_{1} - p_1)^2 = M^2 
\nn\\
\ell_4^2 &=& (- \ell_3 + p_2)^2 = (\ell - p^{\prime}_{2})^2 = 0 
\eea
where $p_1, p_1^{\prime}$ ($p_2, p_2^{\prime}$) are the momenta associated with external $\mu^-$ and $\mu^+$ ($e^-$ and $e^+$), respectively. Using the $3$-particle amplitudes derived above, we find that
\bea
\textrm{Cut-max} &=& \frac{4 e^4}{M^4}
\langle \ell_4 \ell_{I} \rangle \bigl[ \ell_{J} 2^{\prime} \bigr]
\langle \ell_2 \ell^{I} \rangle \bigl[ \ell^{J} 1^{\prime} \bigr]
\langle 1 \ell_3^{K} \rangle \bigl[ \ell_3^{L} \ell_2 \bigr]
\langle 2 \ell_{3 K} \rangle \bigl[ \ell_{3 L} \ell_4 \bigr]
\nn\\
&=& 4 e^4
\langle 1 2 \rangle \langle 1^{\prime} 2^{\prime} \rangle \bigl[ 1^{\prime} 2^{\prime} \bigr]^2 
= - 4 e^4 \frac{s_{12} s_{1^{\prime}2^{\prime}}^2}{\langle 1^{\prime} 2^{\prime} \rangle  \bigl[ 12 \bigr] } .
\eea
As a simple check, one can easily prove that we obtain the same result by resorting to the standard evaluation in terms of polarization sums:
\bea
\textrm{Cut-max} &=& e^4
\langle \ell_4| \gamma^{\mu} |2^{\prime} \bigr]
\langle \ell_2| \gamma^{\nu} |1^{\prime} \bigr]
\left( \eta_{\mu\nu} - \frac{\ell_{\mu} \ell_{\nu}}{M^2} \right)
\langle 1| \gamma^{\alpha} |\ell_2 \bigr]
\langle 2| \gamma^{\beta} |\ell_4 \bigr]
\left( \eta_{\alpha\beta} - \frac{\ell_{3 \alpha} \ell_{3 \beta}}{M^2} \right)
\nn\\
&=& e^4
\left( 2 \langle \ell_4  \ell_2 \rangle \bigl[ 1^{\prime} 2^{\prime} \bigr]
- \frac{ \langle \ell_4| \boldsymbol\ell |2^{\prime} \bigr]\langle \ell_2| \boldsymbol\ell |1^{\prime} \bigr] }{M^2} 
\right)
\left( 2 \langle 12 \rangle \bigl[ \ell_4  \ell_2 \bigr]
- \frac{\langle 1| \boldsymbol\ell_3 |\ell_2 \bigr] \langle 2| \boldsymbol\ell_3 |\ell_4 \bigr]}{M^2} 
\right)
\nn\\
&=& 4 e^4
\langle 1 2 \rangle \langle 1^{\prime} 2^{\prime} \rangle \bigl[ 1^{\prime} 2^{\prime} \bigr]^2 
\eea
where we used the Fierz identity and momentum conservation at each vertex. So we can write that
\beq
A^{1-\textrm{loop}}(e^+ e^- \to \mu^+ \mu^-) =
- 4 e^4 \frac{s_{12} s_{1^{\prime}2^{\prime}}^2}{\langle 1^{\prime} 2^{\prime} \rangle  \bigl[ 12 \bigr] } 
I_{4}(p_{1},p_{1}^{\prime},p_{2},p_{2}^{\prime})
+ \textrm{Triangles} + \textrm{Bubbles} + \textrm{Tadpoles} + {\cal R}
\eeq
for the contribution coming from Merlin particles running inside the loop, and now
\bea
I_{4}(p_{1},p_{1}^{\prime},p_{2},p_{2}^{\prime}) &=& \int \frac{d^{D}\ell}{(2\pi)^D} 
\frac{-i}{[\ell^2 - M^2 - i M \Gamma]} \frac{i}{\left(\ell + p_{1}^{\prime}\right)^2 + i \epsilon}
\nn\\
&\times& \frac{-i}{[\left(\ell + p_{1}^{\prime} + p_{1}\right)^2 - M^2 - i M \Gamma]} 
\frac{i}{\left(\ell - p_{2}^{\prime}\right)^2 + i \epsilon} .
\eea
Observe the change in the overall sign of the Merlin propagators -- as well as in their imaginary parts -- in comparison with the $W$-boson propagators discussed in the previous subsection. Moreover, the presence of $\Gamma$ can be understood along the same lines as in the normal case -- it is important for defining the contour associated with the loop integration but it cannot appear in the final answer obtained after performing the loop integral. Indeed, integrals associated with the Merlin propagators have to be evaluated using the Lee-Wick prescription for integration in the complex $\ell^{0}$ plane so that the Wick rotation remains well defined~{\cite{Lee:1969fy,Cutkosky:1969fq,Coleman:1969xz,Grinstein:08,Grinstein:2008bg}}.

\subsection{Non-local theories}

Previously we have claimed that we can work only with stable particles at the expense of locality. That is, when one is off resonance, a way to deal with the problem of unstable particles is to eliminate them altogether and as a consequence we introduce a non-local description of the problem, albeit one containing only stable modes. Let us briefly discuss this method for the case of light-light scattering. A similar reasoning can also be used for the case of Lee-Wick theories. Since what one has is a non-local interaction, we will discuss the process $\gamma \gamma \to \gamma \gamma$ within a non-local effective theory. We consider a variant of the theory described in Ref.~\cite{Beenakker:00}, i.e., we will consider a non-local scalar QED. This effective context should account for the issues that one will face when dealing with unstable particles off resonance. The non-local interaction between complex scalars and photons is described by the following Lagrangian density
\beq
{\cal L}_{\textrm{NL}} = \phi^{*}(x) \Sigma(x-y) U(x,y) \phi(y)
\eeq
where the non-local coefficient $\Sigma(x-y)$, which plays the role of a scalar self-energy term, is assumed to be a function of the scalar invariant $(x-y)^2$. In addition, the path-ordered exponential $U(x,y)$ is defined as
\beq
U(x,y) = \textrm{P} \exp\left[ -i e \int^{y}_{x} d\omega^{\mu} A_{\mu}(\omega)  \right]
\eeq
where $d\omega^{\mu}$ is the element of integration along a path connecting points $x$ and $y$. The path ordering in the definition of $U(x,y)$ is necessary to maintain the gauge-transformation property of $U(x,y)$ for the non-Abelian case. In any case, the path ordering is not required for the photon case~\cite{Terning:1991yt}. There are also further conditions that should be imposed on the path~\cite{Beenakker:00}. The non-local gauge-boson-scalar-scalar vertex can be derived in the standard way and the result is (assuming that the non-local coefficient has analyticity properties resembling standard self-energy functions)
\beq
i \Gamma^{\mu}(q,p,-p^{\prime}) = i e (2 p + q)^{\mu} {\cal S}(p,p')
(2 \pi)^4 \delta(q + p -p^{\prime})
\eeq
where
\beq
{\cal S}(p^2,p^{\prime 2}) = \frac{ \Bigl[ \Sigma(p^{\prime 2}) - \Sigma(p^{2}) \Bigr] }{p^{\prime 2} - p^{2}},
\eeq
$\Sigma(p^{2})$ being the Fourier transform of the self-energy $\Sigma$. It is easy to verify that, for the full vertex (containing also the local part, which is not of interest to us here), the Ward identity for dressed scalar propagators is respected~\cite{Beenakker:00}. As discussed above, any poles coming from non-local vertices should have zero residues. It is easy to see that this is the case. Furthermore, notice also that
\beq
\lim_{p^2 \to p^{\prime 2}} {\cal S}(p^2,p^{\prime 2}) 
= \frac{\partial \Sigma(p^{\prime 2})}{\partial p^{\prime 2}}  
\equiv F(p^{\prime 2}) .
\eeq
This implies the following three-particle amplitudes involving two complex scalars and one photon:
\bea
A_{3}^{\textrm{tree}}[p,q^{+},p^{\prime}] &=& 2 e F(m^2) \frac{\langle \xi | p | q \bigr]}{\langle \xi q \rangle}
\nn\\
A_{3}^{\textrm{tree}}[p,q^{-},p^{\prime}] &=& 2 e F(m^2) \frac{\langle q | p | \xi \bigr]}{\bigl[ q \xi \bigr]} 
\eea
where $m$ is the mass of the scalars. Observe that this is a gauge-invariant amplitude. Since the other gauge interactions are determined from the condition of off-shell gauge invariance, they should not comprise any new on-shell information. Hence calculation of higher-point tree-level amplitudes may proceed via the usual BCFW recursion relations~\footnote{We recall that a $4$-scalar interaction is also possible in scalar QED and this interaction provides independent gauge-invariant data. As a consequence, the amplitude calculated with the BCFW recursion relation is the one associated with choosing the scalar self-coupling constant to be proportional to $e^2$. For more details regarding this issue in scalar QED -- which is independent of the addition of the non-local interaction -- see Ref.~\cite{Elvang:15}.}.

The one-loop contribution to the process $\gamma \gamma \to \gamma \gamma$ proceeding through scalar loops with the above non-local interaction can be calculated {in the same way as in the previous case} with the $W$ boson. For instance, for the all-plus helicity amplitude one finds that
\beq
A^{1-\textrm{loop}}_{4}(++++) = - 32 e^4 m^4 [F(m^2)]^4 \frac{ s_{12} s_{23} }
{\langle 1 2 \rangle \langle 23 \rangle \langle 34 \rangle \langle 41 \rangle}
I_{4}(p_{1},p_{2},p_{3},p_{4}) + \textrm{Perm.}  + {\cal R}
\eeq
where an extra factor of two was taken into account due to the fact that there is a complex scalar propagating in the loop.

\section{Summary}

{Here we have discussed the use of unitarity methods in field theories containing resonances. We have shown, through the detailed assessment of three physical situations, how the technique can still be put in practice to such theories at one-loop. Our purpose was to provide an one-loop proof of the validity of unitarity-based methods for unstable particles as a consequence of unitarity itself. That is, generically speaking, if unitarity is satisfied by the inclusion of only stable states in unitarity sums, this implies that, for the unitarity method, one must sum over only the asymptotic states of the theory in Eq.~(\ref{2}). In the complex-mass scheme, this means that, to leading order and close to the resonance region, the cut of the unstable propagator proceeds through the cut of the loop of stable particles. On the other hand, in the NWA, the cut taken through the unstable particle (setting its width to zero) recovers the same result as a cut through the stable decay products. This is the basic requirement for a four-dimensional amplitude involving unstable particles to have cut-constructible parts. Without it, unitarity cuts could not enable one to establish a relationship between the pole structure of the integrand and the branch-cut structure of the loop integral.}

{Our aim here was not to devise a complete account of all the aspects of the method for unstable particles. Indeed, even though the proof of applicability of unitarity-based methods might be extended to higher orders in perturbation theory, in the present study we have limited ourselves to one-loop order. This is because unitarity cuts provide useful information that can be used in the most efficient way when a complete basis of integrals is known, and this is somewhat straightforward at one loop; the set of master integrals necessary to perform the reduction of generic one-loop tensor integrals is well known, as discussed above. On the other hand, at leading order and close to the resonance region, the CMS cut propagator of the unstable particle turns into a nascent delta function, reproducing the stable particle result, and hence we are allowed to associate the outcome with physical particles carrying positive energy. Such observations allows one to prove Eq.~(\ref{2}) in a straightforward way. In turn, through the examination of simple one-loop examples, we demonstrated explicitly how powerful the method still is when constructing one-loop amplitudes with unstable particles. Further work is recommended to better understand the application of such methods to these theories. One should establish the validity of the method to higher loops: In principle, the proof to higher orders proceeds in much the same way -- however, in this case other methods (such as integration-by-parts techniques) should also be employed as integral reduction of the loop integral in terms of a set of master integrals is no longer straightforward and the cuts of internal propagators become more involved.}

We have tried to fill this aforementioned gap in the literature of unitarity methods with this primary exploration, and we believe that our study can be useful in the investigations of the Standard Model Effective Field theory (SMEFT)~\cite{Brivio:2017vri}, or in the Higgs Effective Field Theory (HEFT)~\cite{Krause:16}. Indeed, the calculation of loop amplitudes of massive on-shell SMEFT amplitudes focusing on the electroweak sector will obviously involve internal resonances, and an understanding of the unitarity method as a framework to tackle this computation would be most welcome. On the other hand, loop amplitudes of higher-derivative theories also contains resonances, the Merlin modes, and now a careful treatment of those within unitarity methods is available. We believe this will have an important impact on the evaluation of amplitudes of quadratic gravity, a promising conservative ultraviolet completion of quantum gravity. This would indeed be interesting to investigate, and we hope to explore this calculation in subsequent works~\cite{Menezes:2021dyp,Menezes:2022jow}.

\section*{Acknowledgements} 

We thank John F. Donoghue for useful discussions and for collaborations on related topics. This work has been partially supported by Conselho Nacional de Desenvolvimento Cient\'ifico e Tecnol\'ogico - CNPq under grant 310291/2018-6 and Funda\c{c}\~ao Carlos Chagas Filho de Amparo \`a Pesquisa do Estado do Rio de Janeiro - FAPERJ under grant E-26/202.725/2018.

\section*{Appendix -- Quick review of the spinor-helicity formalism}

Here we quickly review some basic aspects concerning the spinor-helicity formalism. For a more detailed discussion concerning massless particles, {we refer the reader to} Refs.~\cite{Schwartz:13,Elvang:15,Henn:2014yza}.

We use Pauli matrices when representing lightlike momenta as bispinors: 
\bea
p^{\alpha \dot{\alpha}} = \sigma^{\alpha \dot{\alpha}}_{\mu} p^{\mu} =
\begin{pmatrix}
p^0 - p^3 && - p^1 + i p^2 \\
- p^1 - i p^2 && p^0 + p^3 
\end{pmatrix} .
\eea
Since $p^{\mu}$ is lightlike, $\det p^{\alpha \dot{\alpha}} = 0$ and hence $p^{\alpha \dot{\alpha}}$ has rank $1$. This means that it can be written as an outer product of helicity spinors:
\beq
p^{\alpha \dot{\alpha}} = \lambda^{\alpha} \tilde{\lambda}^{\dot{\alpha}} .
\eeq
Inner products between helicity spinors {are carried out} with the 2-dimensional Levi-Civita symbol as well as the raising and lowering of spinor indices. Helicity spinors can also be represented as
\bea
\lambda^{\alpha} &=& | p \rangle,  \,\,\, \lambda_{\alpha} = \langle p |
\nn\\
\tilde{\lambda}^{\dot{\alpha}} &=& \bigl [p |, \,\,\, \tilde{\lambda}_{\dot{\alpha}} = | p \bigr] .
\eea
Hence
\beq
p^{\alpha \dot{\alpha}} = | p \rangle \bigl[p |, \,\,\,
p_{\dot{\alpha} \alpha} =  | p \bigr]  \langle p | .
\eeq
We take all momenta incoming, so conservation of momentum reads $\sum_{i} p^{\mu}_{i} = 0$. Since
$$
p^{\mu} = \frac{1}{2} \sigma^{\mu \alpha \dot{\alpha}} p_{\dot{\alpha} \alpha}
$$
conservation of momentum in terms of helicity spinors reads
\beq
\sum_{j=1}^{n} | j \rangle \bigl [j | = 0.
\eeq
In order to write photon polarizations in terms of helicity spinors, one introduces a reference light-like momentum $r^{\mu}$. Except for the fact that it must not be aligned with the associated momentum of the particle, $r^{\mu}$ is arbitrary. For the two physical polarizations, one finds
\bea
\bigl[ \epsilon^{-}_{p}(r) \bigr]^{\alpha \dot{\alpha}} &=& \sqrt{2} \, \frac{ | p \rangle \bigl[r |}{ \bigl[ p r \bigr] },
\,\,\,
\bigl[ \epsilon^{-}_{p}(r) \bigr]_{\dot{\alpha}\alpha } = \sqrt{2} \, \frac{ | r \bigr]  \langle p |}{ \bigl[ p r \bigr] }
\nn\\
\bigl[ \epsilon^{+}_{p}(r) \bigr]^{\alpha \dot{\alpha}} &=& \sqrt{2} \, \frac{ | r \rangle \bigl[p |}{ \langle r p \rangle },
\,\,\,
\bigl[ \epsilon^{+}_{p}(r) \bigr]_{\dot{\alpha}\alpha } = \sqrt{2} \, \frac{ | p \big] \langle r |}{ \langle r p \rangle }
\label{gpolarization}
\eea
where
$$
\bigl[ \epsilon^{\pm}_{p}(r) \bigr]^{\alpha \dot{\alpha}} = \sigma^{\alpha \dot{\alpha}}_{\mu} 
\epsilon^{\mu}_{\pm}(p;r) .
$$
and
\bea
\epsilon^{\mu}_{-}(p;r) &=& \frac{1}{\sqrt{2}} \, \frac{ \langle p | \gamma^{\mu} | r \bigr]  }{ \bigl[ p r \bigr] }
\nn\\
\epsilon^{\mu}_{+}(p;r) &=& \frac{1}{\sqrt{2}} \, \frac{ \langle r | \gamma^{\mu} | p \big] }{ \langle r p \rangle } .
\eea
The Ward identity in QED is assured by the freedom of choice of reference momentum.

Left-handed ($h= - 1/2$) and right-handed ($h= + 1/2$) Dirac spinors can be written as
\bea
| p \rangle &=& 
\begin{pmatrix}
\lambda^{\alpha} \\
0
\end{pmatrix},
\,\,\,
| p \bigr] = 
\begin{pmatrix}
0 \\
\tilde{\lambda}_{\dot{\alpha}}
\end{pmatrix}
\,\,\, (\textrm{incoming fermion, outgoing anti-fermion})
\nn\\
\langle p | &=& 
\begin{pmatrix}
\lambda_{\alpha} && 0
\end{pmatrix},
\,\,\,
\bigl[ p | = 
\begin{pmatrix}
0 && \tilde{\lambda}^{\dot{\alpha}}
\end{pmatrix} 
\,\,\, (\textrm{outgoing fermion, incoming anti-fermion}).
\eea
The gamma matrices in the Weyl basis {read}
\beq
\gamma^{\mu}_{\alpha \dot{\alpha}} = 
\begin{pmatrix}
0 && \sigma^{\mu\alpha \dot{\alpha}} \\
\bar{\sigma}^{\mu}_{\dot{\alpha} \alpha} && 0
\end{pmatrix}
\eeq
where $\sigma^{\mu} = ({\bf 1} \,\,\, \boldsymbol\sigma)$ and 
$\bar{\sigma}^{\mu} = ({\bf 1} \,\, -\boldsymbol\sigma)$. 

Let us briefly discuss the formalism for massive particles that we used in this work~\cite{Arkani-Hamed:17,Shadmi:19,Chung:19,Aoude:19,Durieux:20}. This is obtained by noting that $\det p^{\alpha \dot{\alpha}} = m^2$ in the massive case and now $p^{\alpha \dot{\alpha}}$ has rank $2$. So it can be written as the sum of two rank-one matrices: 
\beq
p^{\alpha \dot{\alpha}} = \lambda^{\alpha\, I} \tilde{\lambda}^{\dot{\alpha}}_{I} .
\eeq
The index $I=1,2$ indicates a doublet of the $SU(2)$ little group. Since $\det p^{\alpha \dot{\alpha}} = \det\lambda \det\tilde{\lambda} = m^2$, we simply take $\det\lambda = \det\tilde{\lambda} = m$. Just like spinor indices, the little group indices are raised and lowered by the $SU(2)$-invariant tensor 
$\varepsilon^{IJ}, \varepsilon_{IJ}$. It implies that
\bea
p^{\alpha \dot{\alpha}} &=& \lambda^{\alpha\, I} \tilde{\lambda}^{\dot{\alpha}}_{I} 
= - \lambda^{\alpha}_{I} \tilde{\lambda}^{\dot{\alpha}\,I}
= | p^{I} \rangle \bigl[p_{I} |
\nn\\
p_{\dot{\alpha} \alpha} &=& - \tilde{\lambda}_{\dot{\alpha}}^{I} \lambda_{\alpha\, I} 
= \tilde{\lambda}_{\dot{\alpha} \, I}  \lambda_{\alpha}^{I} 
= - | p^{I} \bigr]  \langle p_{I} |.
\eea
By definition, the massive spinor helicity variables satisfy
\beq
p^{\alpha \dot{\alpha}} \tilde{\lambda}_{\dot{\alpha}}^{I} = m \lambda^{\alpha\, I},
\,\,\,
p_{\dot{\alpha} \alpha} \lambda^{\alpha\,I}  = m \tilde{\lambda}_{\dot{\alpha}}^I .
\eeq
Comparing this with the usual Dirac equations of motion {one is led to the natural identifications} for the Dirac spinors:
\bea
u^{I}(p) &=& 
\begin{pmatrix}
\lambda^{\alpha\, I} \\
\tilde{\lambda}_{\dot{\alpha}}^{I}
\end{pmatrix}
\nn\\
v^{I}(p) &=& 
\begin{pmatrix}
\lambda^{\alpha\, I} \\
- \tilde{\lambda}_{\dot{\alpha}}^{I}
\end{pmatrix}
\eea
and similarly for the conjugate spinors
\bea
\bar{u}_{I}(p) &=& 
\begin{pmatrix}
- \lambda_{\alpha\, I} &&
\tilde{\lambda}^{\dot{\alpha}}_{I}
\end{pmatrix}
\nn\\
\bar{v}_{I}(p) &=& 
\begin{pmatrix}
\lambda_{\alpha\, I} &&
 \tilde{\lambda}^{\dot{\alpha}}_{I}
\end{pmatrix} .
\eea
Since the massive spinor bilinears satisfy
\bea
\langle \lambda^{I} \lambda_{J} \rangle &=& m \delta^{I}_{J},
\,\,\,
\langle \lambda^{I} \lambda^{J} \rangle = - m \varepsilon^{IJ},
\,\,\,
\langle \lambda_{I} \lambda_{J} \rangle = m \varepsilon_{IJ}
\nn\\
\bigl[ \tilde{\lambda}^{I} \tilde{\lambda}_{J} \bigr] &=& -m \delta^{I}_{J},
\,\,\,
\bigl[ \tilde{\lambda}^{I} \tilde{\lambda}^{J} \bigr] = m \varepsilon^{IJ},
\,\,\,
\bigl[ \tilde{\lambda}_{I} \tilde{\lambda}_{J} \bigr] = - m \varepsilon_{IJ} 
\nn\\
\lambda^{I\,\alpha} \lambda_{I\, \beta} &=&  |\lambda^{I} \rangle \langle \lambda_{I} |
= -m \delta^{\alpha}_{\beta},
\,\,\,
\tilde{\lambda}^{I}_{\dot{\alpha}} \tilde{\lambda}_{I}^{\dot{\beta}} =  
|\lambda^{I} \bigr] \bigl[ \lambda_{I} |
= m \delta_{\dot{\alpha}}^{\dot{\beta}}
\label{relations}
\eea
it is easy to see that the Dirac spinors obey the usual spin sums. Let us introduce a bold notation to indicate symmetric compositions of the $SU(2)$ little-group indices of massive spinors. One has that 
\bea
p^{\alpha \dot{\alpha}} &=& | {\bf p} \rangle \bigl[ {\bf p} |
\nn\\
p_{\dot{\alpha} \alpha} &=& - | {\bf p} \bigr]  \langle {\bf p} |
\eea
and the Dirac equation can be rewritten as
\bea
p^{\alpha \dot{\alpha}}  | {\bf p} \bigr] &=& m | {\bf p} \rangle
\nn\\
p_{\dot{\alpha} \alpha} | {\bf p} \rangle &=& m | {\bf p} \bigr] 
\nn\\
\bigl[ {\bf p} | p_{\dot{\alpha} \alpha} &=& - m \langle {\bf p} |
\nn\\
\langle {\bf p}|  p^{\alpha \dot{\alpha}} &=& -m \bigl[ {\bf p} | .
\eea
In addition:
\bea
\langle {\bf 3}1 \rangle \langle {\bf 3}2 \rangle &=&
\langle 3^I 1 \rangle \langle 3^I 2 \rangle
\,\,\, (I=J)
\nn\\
\langle {\bf 3}1 \rangle \langle {\bf 3}2 \rangle &=&
\frac{1}{\sqrt{2}} \Bigl( \langle 3^I 1 \rangle \langle 3^J 2 \rangle 
+ \langle 3^J 1 \rangle \langle 3^I 2 \rangle \Bigr)
\,\,\, (I \neq J) .
\eea
We can alternatively write the massive momentum as
\beq
p^{\mu} = k^{\mu} + q^{\mu}
\eeq
where $k^2 = q^2 = 0$ and $p^2 = 2 k \cdot q = \langle kq \rangle \bigl[ qk \bigr] = m^2$. In terms of bispinors:
\bea
p^{\alpha \dot{\alpha}} &=& k^{\alpha \dot{\alpha}} + q^{\alpha \dot{\alpha}}
\nn\\
p_{\dot{\alpha} \alpha} &=& k_{\dot{\alpha} \alpha} + q_{\dot{\alpha} \alpha} .
\eea
One has the following identifications
\bea
| p^1 \rangle &=& | q \rangle,
\,\,\,
| p_1 \bigr] = | q \bigr]
\nn\\
| p^2 \rangle &=& | k \rangle,
\,\,\,
| p_2 \bigr] = | k \bigr]
\nn\\
| p^1 \bigr] &=& | k \bigr],
\,\,\,
| p_1 \rangle = - | k \rangle
\nn\\
| p^2 \bigr] &=& - | q \bigr],
\,\,\,
| p_2 \rangle = | q \rangle .
\eea
Also
\bea
\langle kq \rangle &=&  \bigl[ qk \bigr] = m
\nn\\
p^{\alpha \dot{\alpha}}  | k \bigr] &=& m | q \rangle
\nn\\
p_{\dot{\alpha} \alpha} | k \rangle &=& - m | q \bigr] 
\nn\\
p^{\alpha \dot{\alpha}}  | q \bigr] &=& - m | k \rangle
\nn\\
p_{\dot{\alpha} \alpha} | q \rangle &=& m | k \bigr] .
\eea
The polarization vector of a massive vector boson of momentum $p$ and mass $m$ is given by
\beq
\epsilon^{IJ}_{\mu} = \frac{1}{\sqrt{2}} \, \frac{ \langle {\bf p} | \sigma_{\mu} | {\bf p} \bigr]  }{m}
\eeq
or, in terms of bispinors:
\beq
\bigl[ \epsilon^{IJ} \bigr]^{\alpha \dot{\alpha}} = \sqrt{2} \, \frac{ | {\bf p} \rangle \bigl[ {\bf p} |}{m} .
\eeq
There is an implicit symmetrization on $SU(2)$ indices. These polarizations correspond to transverse 
and longitudinal modes:
\bea
\epsilon^{+}_{\mu} &=& \epsilon^{11}_{\mu} 
= \frac{1}{\sqrt{2}} \, \frac{ \langle p^1 | \sigma_{\mu} | p^1 \bigr]  }{m}
\nn\\
\epsilon^{-}_{\mu} &=& \epsilon^{22}_{\mu} 
= \frac{1}{\sqrt{2}} \, \frac{ \langle p^2 | \sigma_{\mu} | p^2 \bigr]  }{m}
\nn\\
\epsilon^{0}_{\mu} &=& \epsilon^{12}_{\mu} = \epsilon^{21}_{\mu}
= \frac{1}{2} \, \frac{ \langle p^1 | \sigma_{\mu} | p^2 \bigr] 
+ \langle p^2 | \sigma_{\mu} | p^1 \bigr]  }{m} .
\eea
These massive polarization vectors satisfy the traditional normalization for the vector boson.

Under little group scaling, one finds that
\beq
| p \rangle \to z | p \rangle,
\,\,\,
| p \bigr] \to z^{-1} | p \bigr]
\eeq
for massless spinors, whereas for massive spinors we have the following $SL(2)$ transformation:
\beq
\lambda^{I} \to W^{I}_{J} \lambda^{J}
\,\,\,
\tilde{\lambda}_{I} \to (W^{-1})_{I}^{J} \tilde{\lambda}_{J} .
\eeq

\end{document}